\documentclass[%
 reprint,
superscriptaddress,
 amsmath,amssymb,
 aps,
floatfix,
]{revtex4-2}

\usepackage{graphicx}
\usepackage{dcolumn}
\usepackage{bm}
\usepackage{makecell}

\begin{document}

\preprint{APS/123-QED}

\title{Searching for resonance states in $^{22}$Ne($p,\gamma$)$^{23}$Na}

\author{D. P. Carrasco-Rojas}
\email{dpcarrascor@miners.utep.edu}
\affiliation{Department of Physics, The University of Texas at El Paso, El Paso, TX 79968-0515, USA}
\author{M. Williams}%

\altaffiliation[Current address: ]{Department of Physics, University of Surrey, Guildford GU2 7XH, United Kingdom}
\email{matthew.williams@surrey.ac.uk}
\affiliation{TRIUMF, Vancouver, BC V6T 2A3, Canada}%
\affiliation{Department of Physics, University of York, Heslington, York, YO10 5DD, United Kingdom}

\author{P. Adsley}
\email{padsley@tamu.edu}
\affiliation{Department of Physics and Astronomy, Texas A\&M University, College Station, Texas 77843-4242, USA}
\affiliation{Cyclotron Institute, Texas A\&M University, College Station, Texas 77843-3636, USA}
\affiliation{iThemba Laboratory for Accelerator Based Sciences, Somerset West 7129, South Africa}
\affiliation{School of Physics, University of the Witwatersrand, Johannesburg 2050, South Africa}

\author{L. Lamia}
\affiliation{Laboratori Nazionali del Sud - Istituto Nazionale di Fisica Nucleare, Via Santa Sofia 62, 95123 Catania, Italy}
\affiliation{Dipartimento di Fisica e Astronomia ``E.Majorana", Universit\`a di Catania, Italy}
\affiliation{Centro Siciliano di Fisica Nucleare e Struttura della Materia (CSFNSM), Catania, Italy}

\author{B. Bastin}
\affiliation{GANIL, CEA/DRF-CNRS/IN2P3, Bvd Henri Becquerel, 14076 Caen, France}

\author{T. Faestermann}
\affiliation{Physik Department E12, Technische Universit\"at M\"unchen, D-85748 Garching, Germany}

\author{C. Foug\`eres}
\affiliation{GANIL, CEA/DRF-CNRS/IN2P3, Bvd Henri Becquerel, 14076 Caen, France}

\author{F. Hammache}
\affiliation{Universit\'e Paris-Saclay, CNRS/IN2P3, IJCLab, 91405 Orsay, France}

\author{D. S. Harrouz}
\affiliation{Universit\'e Paris-Saclay, CNRS/IN2P3, IJCLab, 91405 Orsay, France}

\author{R. Hertenberger}
\affiliation{Fakult\"at f\"ur Physik, Ludwig-Maximilians-Universit\"at M\"unchen, D-85748 Garching, Germany}

\author{M. La Cognata}
\affiliation{Laboratori Nazionali del Sud - Istituto Nazionale di Fisica Nucleare, Via Santa Sofia 62, 95123 Catania, Italy}

\author{A. Meyer}
\affiliation{Universit\'e Paris-Saclay, CNRS/IN2P3, IJCLab, 91405 Orsay, France}

\author{F. de Oliveira Santos}
\affiliation{GANIL, CEA/DRF-CNRS/IN2P3, Bvd Henri Becquerel, 14076 Caen, France}

\author{S. Palmerini}
\affiliation{Dipartimento di Fisica e Geologia, Universit\`a degli Studi di Perugia, via A. Pascoli s/n, 06125 Perugia, Italy}
\affiliation{Istituto Nazionale di Fisica Nucleare - Sezione di Perugia, via A. Pascoli s/n, 06125 Perugia, Italy}

\author{R. G. Pizzone}
\affiliation{Laboratori Nazionali del Sud - Istituto Nazionale di Fisica Nucleare, Via Santa Sofia 62, 95123 Catania, Italy}

\author{S. Romano}
\affiliation{Laboratori Nazionali del Sud - Istituto Nazionale di Fisica Nucleare, Via Santa Sofia 62, 95123 Catania, Italy}

\author{N. de S\'er\'eville}
\affiliation{Universit\'e Paris-Saclay, CNRS/IN2P3, IJCLab, 91405 Orsay, France}

\author{A. Tumino}
\affiliation{Laboratori Nazionali del Sud - Istituto Nazionale di Fisica Nucleare, Via Santa Sofia 62, 95123 Catania, Italy}
\affiliation{Facolt\`a di Ingegneria e Architettura, Universit\`a degli Studi di Enna ``Kore'', Cittadella Universitaria, 94100 Enna, Italy
}

\author{H.-F. Wirth}
\affiliation{Fakult\"at f\"ur Physik, Ludwig-Maximilians-Universit\"at M\"unchen, D-85748 Garching, Germany}

\date{\today}

\begin{abstract}
\begin{description}
\item[Background] Globular clusters show strong correlations between different elements, such as the well-known sodium-oxygen anticorrelation. One of the main sources of uncertainty in this anticorrelation is the $^{22}$Ne($p,\gamma$)$^{23}$Na reaction rate, due to the possible influence of an unobserved resonance state at $E_\mathrm{x} = 8862$ keV ($E_\mathrm{r, c.m.}  = 68$ keV). The influence of two higher-lying resonance states at $E_\mathrm{x} = 8894$ and $9000$ keV has already been ruled out by direct $^{22}$Ne($p,\gamma$)$^{23}$Na measurements.
\item[Purpose] To study excited states in $^{23}$Na above the proton threshold to determine if the unconfirmed resonance states in $^{23}$Na exist.
\item[Methods] The non-selective proton inelastic scattering reaction at low energies was used to search for excited states in $^{23}$Na above the proton threshold. Protons scattered from various targets were momentum-analysed in the Q3D magnetic spectrograph at the Maier-Leibnitz Laboratorium, Munich, Germany.
\item[Results] The resonance states previously reported at $E_\mathrm{x} = 8862$, $8894$ and $9000$ keV in other experiments were not observed in the present experiment at any angle. This result, combined with other non-observations of these resonance states in most other experiments, results in a strong presumption against the existence of these resonance states.
\item[Conclusions] The previously reported resonance states at $E_\mathrm{x} = 8862$, $8894$ and $9000$ keV are unlikely to exist and should be omitted from future evaluations of the $^{22}$Ne($p,\gamma$)$^{23}$Na reaction rates. Indirect studies using low-energy proton inelastic scattering are a simple and yet exceptionally powerful tool in helping to constrain astrophysical reaction rates by providing non-selective information of the excited states of nuclei.
\end{description}
\end{abstract}

\maketitle


\section{\label{sec:level1}Astrophysical Background}


Globular clusters (GCs) are large populations of stars bound together by gravity. Since GCs are considered to be formed and evolved without external factors, they are one of the most important laboratories for understanding stellar evolution. In contrast to what was believed years ago, they are composed of multiple populations of stars \cite{doi:10.1146/annurev-astro-081817-051839,PhysRevLett.115.252501} and are observed to exhibit anomalous abundance trends in elements from C to Al that are not seen in field stars. These abundance anomalies challenge our understanding of how GCs have formed and evolved, with some unknown polluting site or sites \cite{10.1093/mnras/stw3029,Ventura_2001,DAntona,Denissenkov_2003,10.1093/mnras/stt444,10.1093/mnras/stt1877,Prantzos,Decressin,Mink,10.1111/j.1745-3933.2010.00854.x} within the GCs causing significant star-to-star variations. Suggested polluting sites include low-mass stars, AGB and super-AGB stars, massive and supermassive stars, and classical novae based on carbon-oxygen or oxygen-neon white dwarfs (see Refs. \cite{doi:10.1146/annurev-astro-081817-051839,Iliadis_2016} for evaluations of possible sites).

One of the most uncertain abundance patterns is the sodium-oxygen anti-correlation \cite{PhysRevC.102.035801, PhysRevLett.115.252501}. The abundance of $^{23}$Na has a significant impact on this anti-correlation; the $^{22}$Ne($p,\gamma$)$^{23}$Na reaction, found in the NeNa cycle of hydrogen burning, has a large effect on the abundance of $^{23}$Na along with other reactions such as $^{23}$Na($p,\gamma$)$^{24}$Mg, $^{23}$Na($p,\alpha$)$^{20}$Ne \cite{Cristallo_2011,Lucatello_2011,PhysRevC.104.L032801}.

Different facilities have studied the $^{22}$Ne($p,\gamma$)$^{23}$Na reaction and $^{23}$Na resonance states. Direct measurements of resonance strengths within the astrophysical region of interest are challenging. Indirect measurements, such as $\gamma$-ray spectroscopy and single-proton transfer reactions have provided valuable constraints on the reaction rate. Significant uncertainties of up to an order of magnitude in the reaction rate at temperatures relevant to AGB stars and Hot-Bottom Burning ($T<0.1$ GK) remain due to unclear nuclear data; the reaction rate at temperatures relevant to classical novae is rather well constrained \cite{PhysRevC.102.035801,PhysRevLett.121.172701,Iliadis_2002}.

Direct measurements of resonance strengths have extended down to $E_\mathrm{c.m.} = 65$ keV. The direct measurements performed at the Laboratory for Undeground Nuclear Astrophysics \cite{PhysRevLett.115.252501,PhysRevLett.121.172701}, the Laboratory for Experimental Nuclear Astrophysics \cite{PhysRevC.81.055804,PhysRevC.92.035805,PhysRevC.95.015806} and the DRAGON recoil separator at TRIUMF \cite{LENNARZ2020135539,PhysRevC.102.035801} have provided resonance strengths or upper limits of resonance strengths for most known resonance states \cite{PhysRevC.65.015801, MOSS1976429, PhysRevC.87.064301}. However, three resonances corresponding to states reported by Powers {\it et al.} \cite{PhysRevC.4.2030} at $E_\mathrm{x} = 8862$, $8892$ and $9000$ keV ($E_\mathrm{r, c.m.} = 68$, $100$ and $206$ keV, $E_\mathrm{r, lab} = 71$, $105$ and $215$ keV) have not been observed. In fact, the LUNA direct measurements have ruled out the $E_\mathrm{x} = 8892$- and $9000$-keV resonance strengths as being astrophysically important. However, even with the stringent limit of $\omega\gamma_\mathrm{68} < 6 \times 10^{-11}$ eV (at 90\% confidence) on the strength of the $E_\mathrm{r, c.m.} = 68$-keV resonance provided by Ferraro {\it et al.} \cite{PhysRevLett.121.172701}, the $^{22}$Ne($p,\gamma$)$^{23}$Na reaction rate is uncertain by a factor of ten depending on the existence of this state.

This paper reports an experimental study of the $^{23}$Na($p,p^\prime$)$^{23}$Na proton inelastic-scattering reaction with a beam energy of $E_p = 14$ MeV using the Q3D magnetic spectrometer at Munich \cite{LOFFLER19731} to momentum analyse the scattered particles. At this energy, the proton inelastic scattering reaction is rather insensitive to the structure of the excited states enabling a stringent test of the existence of possible resonance states at $E_\mathrm{x} = 8862$, $8892$ and $9000$ keV. Section \ref{sec:reaction} discusses the applicability of the ($p,p^\prime$) reaction as an unselective probe to search for $^{23}$Na states of interest.  Section \ref{sec:experiment} details the experimental methodology and data analysis procedures. Our results are then presented and discussed in Section \ref{sec:results}, with concluding remarks given in Section \ref{sec:conclusions}.

\section{The ($p,p^\prime$) reaction}
\label{sec:reaction}

The ($p,p^\prime$) reaction has a long history of being used to perform simple spectroscopy to identify states at the energy used in this experiment ($E_p = 14$~MeV) and in similar past experiments \cite{MOSS1976413,MOSS1976429,PhysRevC.89.065805,BenamaraThesis,PhysRevC.97.045807}. One of the primary features of this reaction is that it appears to be non-selective, unlike many of the other reactions used for these experiments such as, e.g. $^{22}$Ne($^3\mathrm{He},d$)$^{23}$Na which is sensitive to the $^{22}$Ne$+p$ structure of the $^{23}$Na states. The evidence for the selectivity of the reaction is empirical; the outline of why this reaction is believed to have weak or no selectivity to the structure of the excited states is summarised below. 

Measurements of the $^{27}$Al($p,p^\prime$)$^{27}$Al reaction have with the Orsay SplitPole and the Munich Q3D \cite{PhysRevC.89.065805,BenamaraThesis} observed every known state in $^{27}$Al between the ground state and the $^{23}$Na$+\alpha$ threshold at $S_\alpha = 10091.8(1)$ keV. The Q3D measurement, which will be the subject of a future publication reports angular distributions which are flat and featureless, implying that the reaction proceeds through a compound process. This is possibly due to the outgoing protons having energies ($4.3$ and $5.1$ MeV) just above the Coulomb barrier height ($V_\mathrm{Coulomb} = 3.3$ MeV) meaning that compound processes dominate over direct ones. One result of this is that no information may be obtained about the spins and parities of the states populated since the characteristic diffraction patterns obtained in higher-energy inelastic scattering, dominated by direct processes, are absent.

All known isolated states between the ground state and the $^{23}$Na$+\alpha$ threshold at $S_\alpha = 10091.8(1)$ keV were observed in Ref. \cite{PhysRevC.89.065805}. This is in contrast to the suggestion of Moss and Sherman \cite{MOSS1976413} that there is a strong antiselectivity to $J=\frac{1}{2}$ states in the ($p,p^\prime$) reaction based on the non-observation of a number of levels in that study. The cause of this disagreement is unclear though we note that the studies of Benamara {\it et al.} \cite{PhysRevC.89.065805} and Adsley {\it et al.} \cite{PhysRevC.97.045807} observed a number of additional states in $^{27}$Al (Benamara) and $^{26}$Mg (Adsley) which were not observed by Moss and Sherman \cite{MOSS1976413} and Moss \cite{MOSS1976429}, suggesting that the experimental conditions may have played a role in their conclusion regarding the antiselection to $J=1/2$ states.

Finally, we note that this proton scattering reaction is not a resonance reaction of the form $^{22}$Ne$+p$ and this means that the cross sections for populating $^{23}$Na states are not thought to be proportional to the proton widths for those $^{23}$Na states. This means that the $^{22}$Ne$+p$ widths required for the $^{22}$Ne($p,\gamma$)$^{23}$Na resonance strengths and subsequently reaction rates cannot be deduced from these data.

\section{Experimental details and data analysis}
\label{sec:experiment}
A 14-MeV proton beam from the tandem accelerator at the Maier-Leibnitz Laboratorium was transported to the target of the Q3D spectrograph \cite{LOFFLER19731}. The target used to probe $^{23}$Na states consisted of around 50 $\mu$g/cm$^2$ of NaF on a 20 $\mu$g/cm$^2$ carbon backing. In order to quantify the background from instrumental effects and target contamination, scattering from a carbon foil similar to the backing of the NaF target was also measured. A LiF target was used to identify $^{19}$F states but only for the 70-degree data. 

Reaction products were momentum analyzed by the Munich Q3D and detected at the focal plane in a detector consisting of two gas-proportional counters backed by a plastic scintillator (for details about the focal-plane detector and similar devices see Refs. \cite{WirthPhD}). Data were collected at five scattering angles: 25, 35, 40, 50, and 70 degrees. Data from different angles are used to identify peaks in the focal plane coming from target contaminants. Target contaminants appear to move in excitation energy with changing angle due to different kinematic shifts. This shift technique has been used to identify contaminating states in other reactions such as the $^{26}$Mg($p,p^\prime$)$^{26}$Mg reaction where a small amount of $^{24}$Mg contamination was present in the target and the resulting peaks could be identified and rejected through the kinematic shift \cite{PhysRevC.97.045807}. 

\begin{figure}[htbp]
    \centering
    \includegraphics[width=0.9\columnwidth]{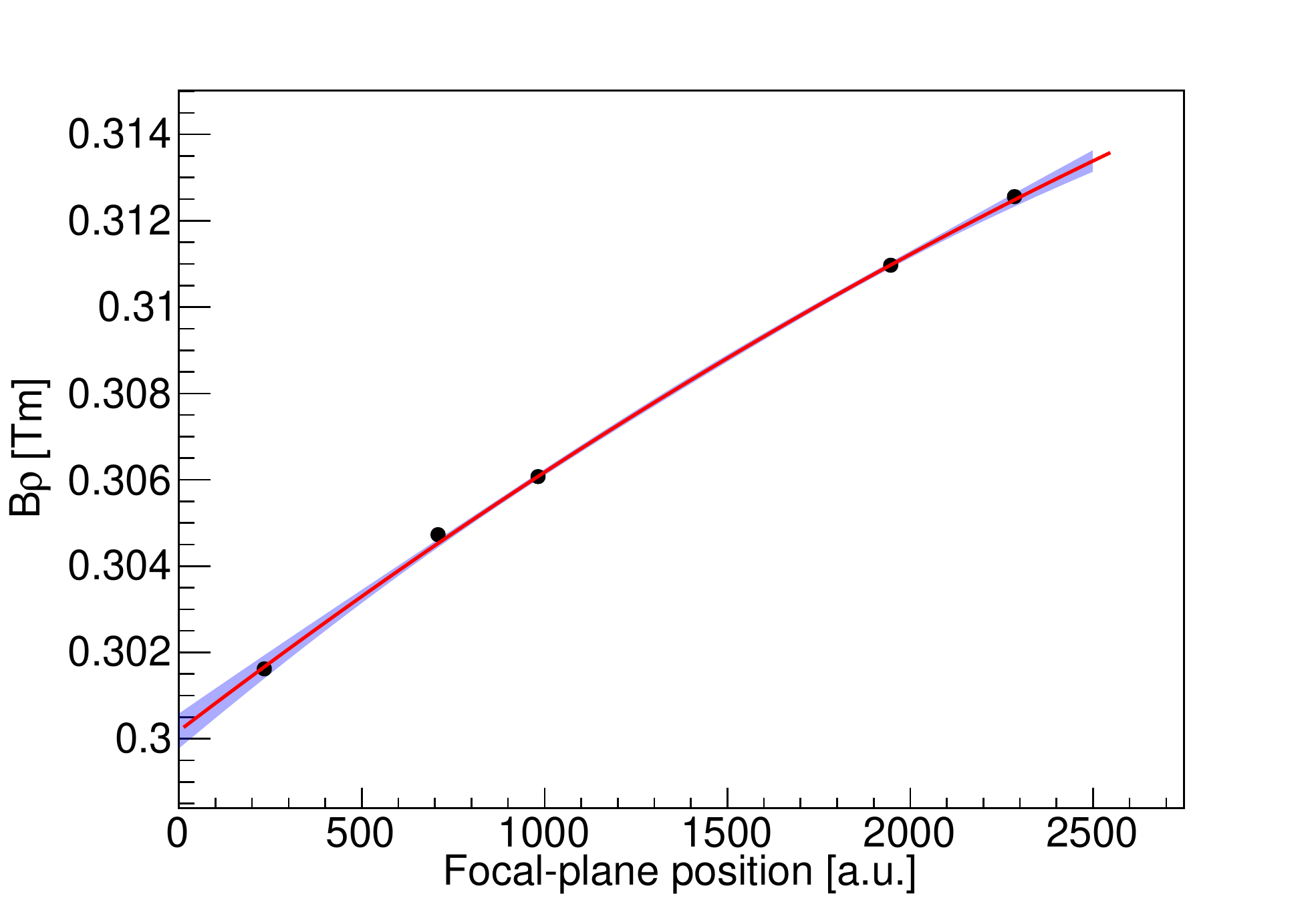}
    \caption{The focal-plane position to magnetic rigidity ($B\rho$) calibration for the data taken at a scattering angle of $\theta_\mathrm{Q3D} = 70$ degrees. The black points correspond to the known levels in $^{19}$F at $E_\mathrm{x} = 8629(4)$, $8864(4)$ and $8953(3)$ keV, and in $^{23}$Na at $E_\mathrm{x} = 8798.7(8)$ and $8945.1(8)$ keV used for the calibration. Both the known uncertainties in the excitation energies listed in the ENSDF \cite{ENSDF} and the fitting uncertainties from the peaks are included in the plot, though the error bars are frequently smaller than the points. The red line represents a best fit with a quadratic function and the blue region shows the $3\sigma$ uncertainty band.}
    \label{fig:70degcalibration}
\end{figure}

\begin{figure}[t]
    \includegraphics[width=0.9\columnwidth]{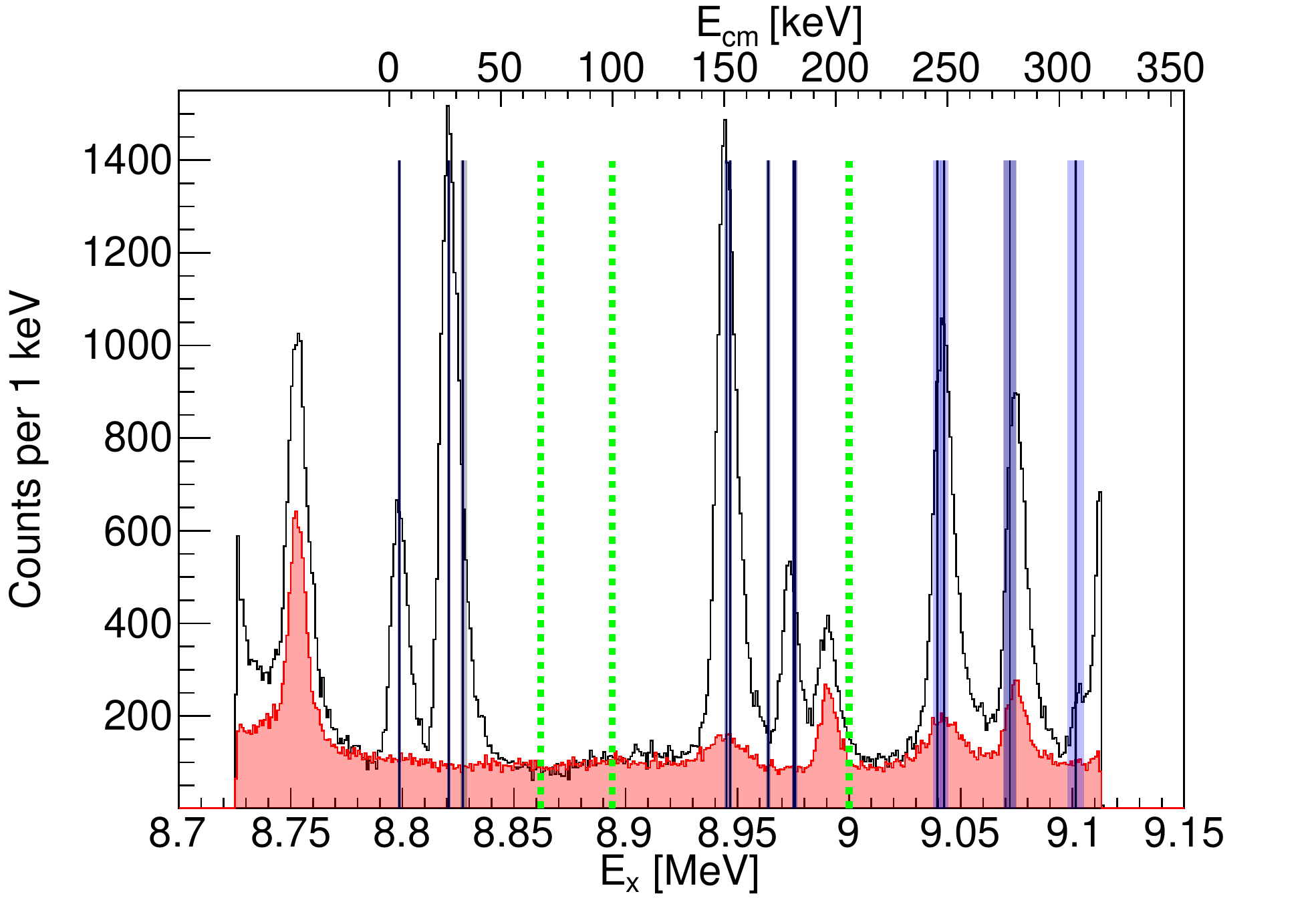}
    \includegraphics[width=0.9\columnwidth]{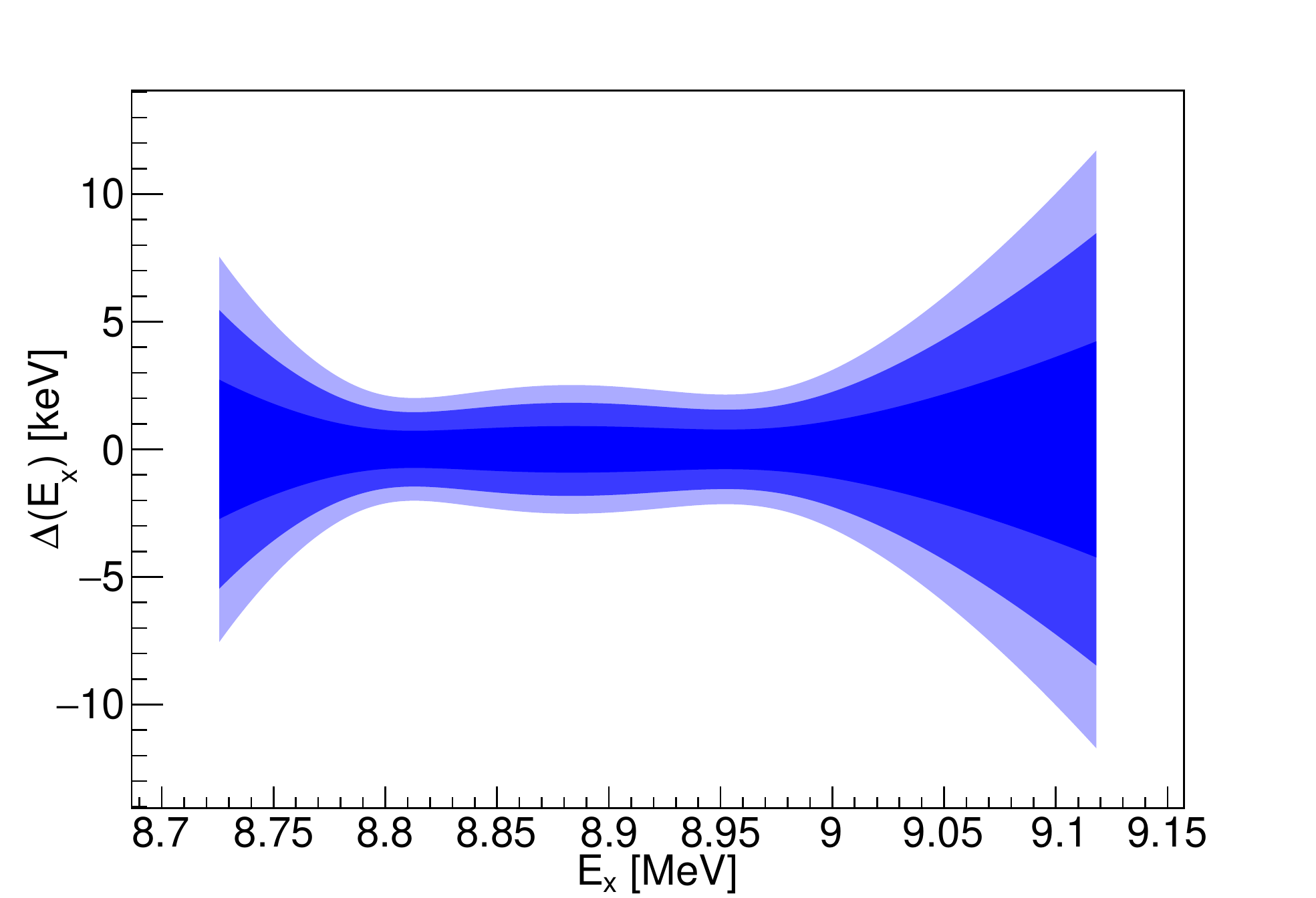}
    \caption{(Top) Excitation-energy spectrum obtained at scattering angle of $\theta_\mathrm{Q3D} = 70$ degrees and assuming $^{23}$Na kinematics. The hollow black-lined spectrum is taken with the NaF target, the solid red-filled spectrum is taken with the LiF target. The solid black vertical lines show the listed energies of levels in $^{23}$Na from the ENSDF \cite{ENSDF} with the associated grey box showing the listed $E_\mathrm{x}$ uncertainty and a blue box showing the $E_\mathrm{x}$ uncertainty from the present measurement. The green dotted vertical lines show the energies of the tentative resonances at $E_\mathrm{x} = 8862$, $8894$ and $9000$ keV from Ref. \cite{PhysRevC.4.2030}. The centre-of-mass energy for the $^{22}$Ne$+p$ colliding system is shown on the top axis. (Bottom) The corresponding $1\sigma$, $2\sigma$ and $3\sigma$ uncertainty bands for the excitation-energy calibration relative to the central value for the 70-degree data.}
    \label{fig:70deg}
\end{figure}

\begin{figure}[t]
    \centering
    \includegraphics[width=0.9\columnwidth]{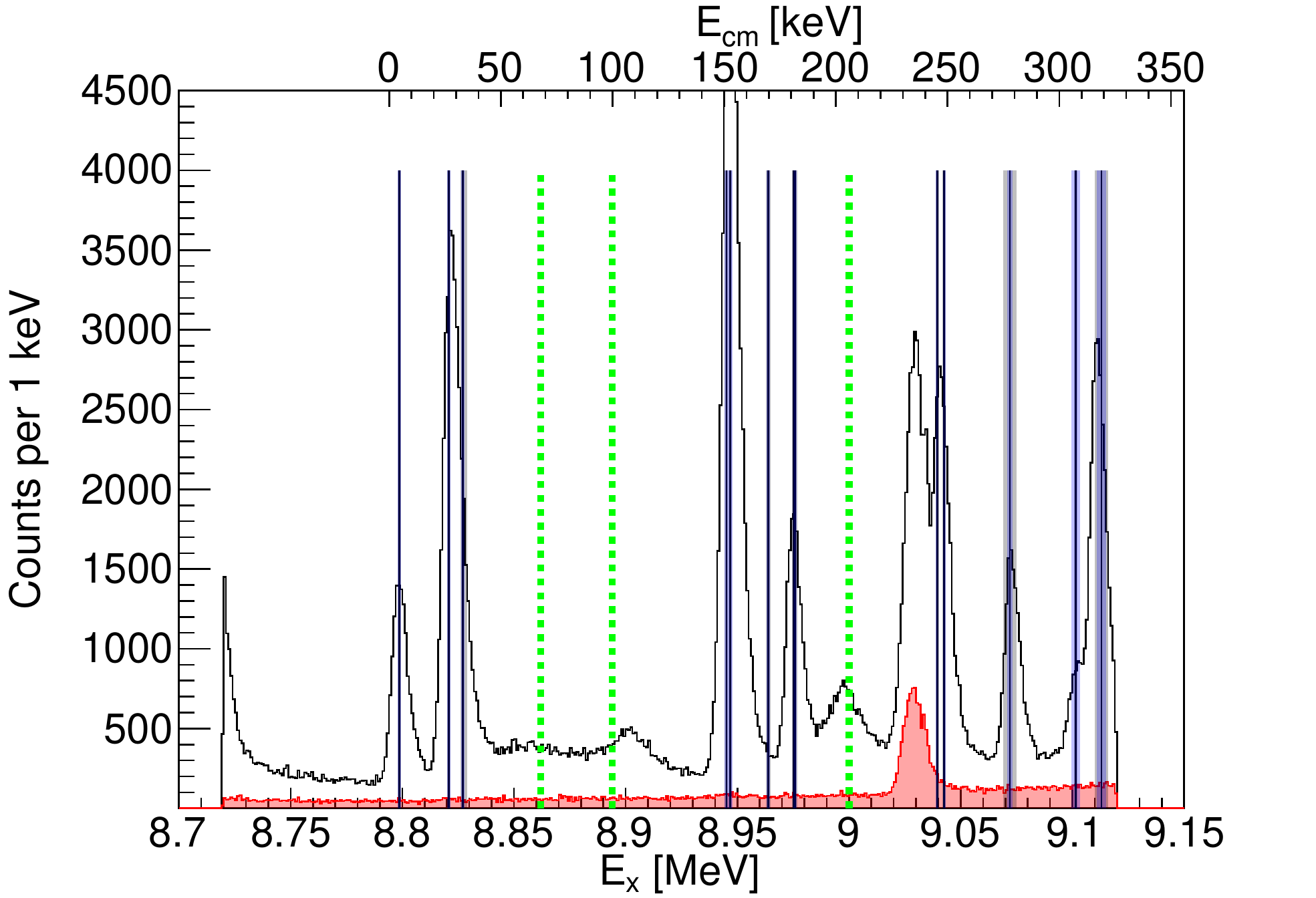}
    \includegraphics[width=0.9\columnwidth]{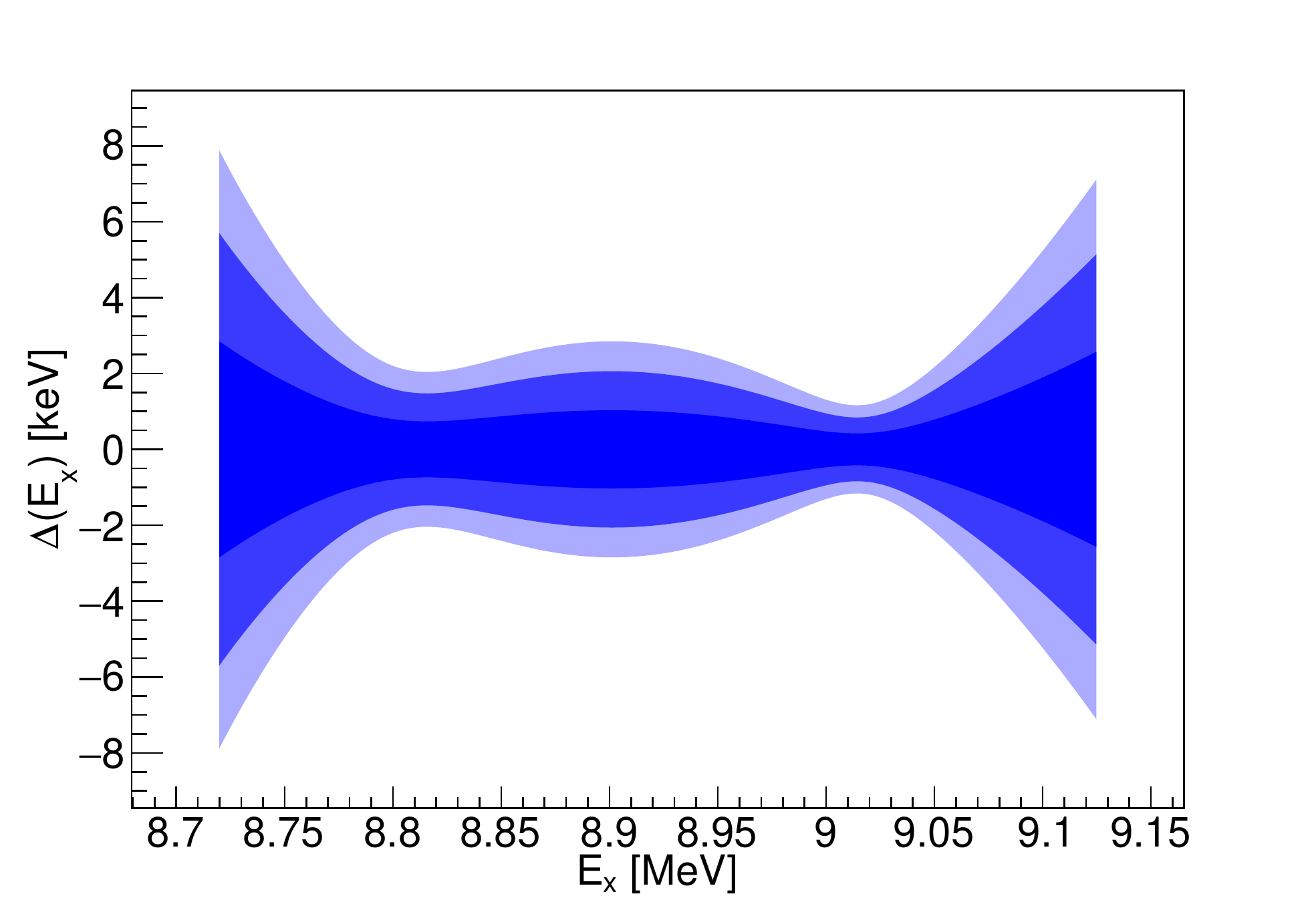}
    \caption{Same as Fig. \ref{fig:70deg} at $\theta_\mathrm{Q3D} = 50$ degrees except that the red filled spectrum is taken with the carbon target; the peak in the spectrum is from a contaminating $^{16}$O state, likely from water or oil on the carbon and NaF targets (or the carbon backing of the NaF target). The spectrum was calibrated using the $^{23}$Na states at $E_\mathrm{x} = 8798.7(8)$, $8975.3(7)$ and $9072(3)$ keV, and the $^{16}$O state at $E_\mathrm{x} = 8871.9(5)$ keV, which is the only peak in the red filled spectrum with the carbon target. The peak at around $E_{\mathrm{x}} = 8900$ keV is an unidentified contaminant since it does not appear at other angles and is broader than the spectrometer resolution, which may be due to the state itself being broad or the kinematics being out of focus.}
    \label{fig:50deg}
\end{figure}

\begin{figure}[t]
    \centering
    \includegraphics[width=0.9\columnwidth]{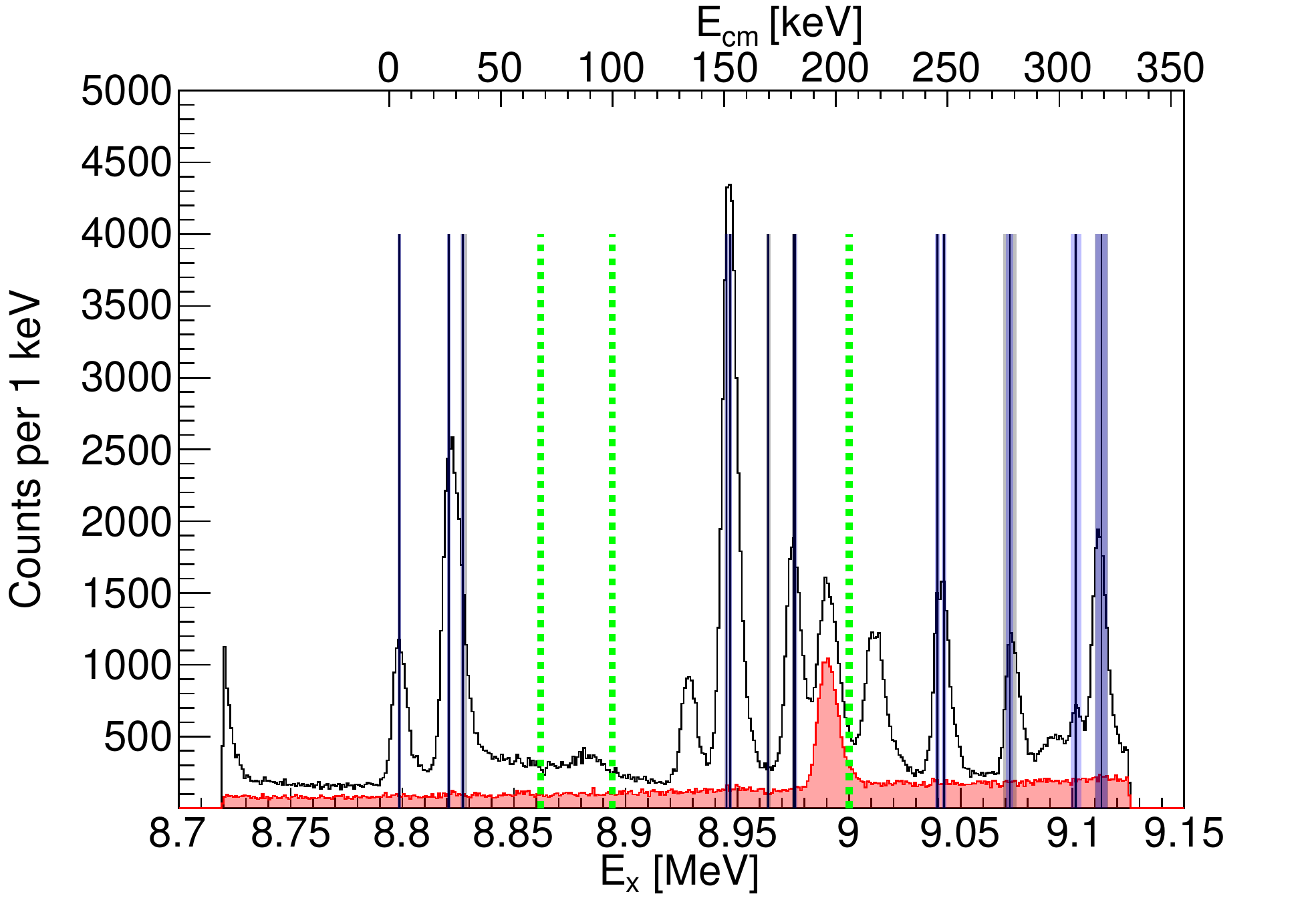}
    \includegraphics[width=0.9\columnwidth]{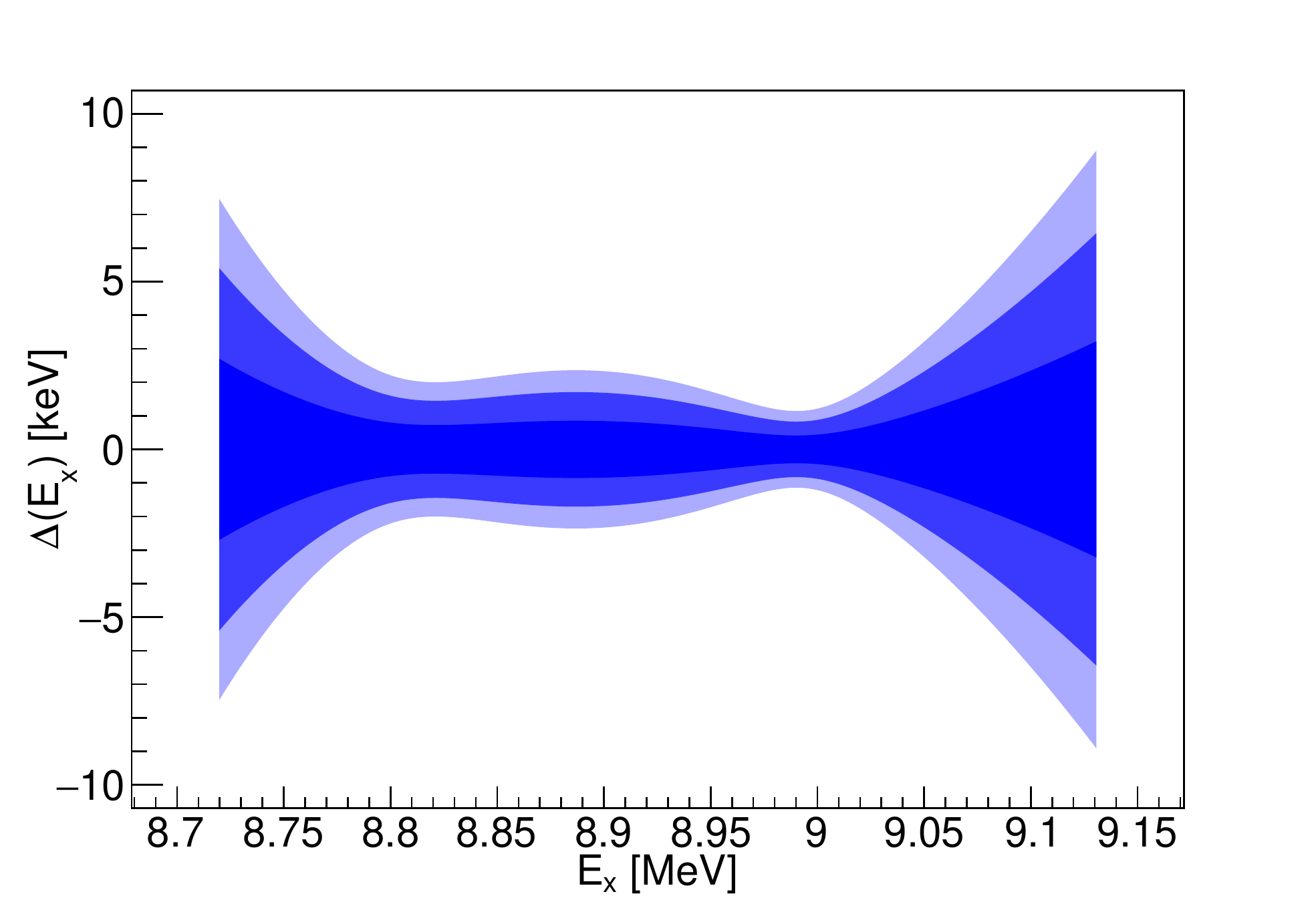}
    \caption{As Fig. \ref{fig:50deg} at $\theta_\mathrm{Q3D} = 40$ degrees. The spectrum was calibrated using the $^{23}$Na states at $E_\mathrm{x} = 8798.7(8)$, $9072(3)$ and $9113(3)$ keV, the $^{19}$F state at $E_\mathrm{x} = 8864(4)$ keV, and the $^{16}$O state at $E_\mathrm{x} = 8871.9(5)$ keV.}
    \label{fig:40deg}
\end{figure}

\begin{figure}[t]
    \centering
    \includegraphics[width=0.9\columnwidth]{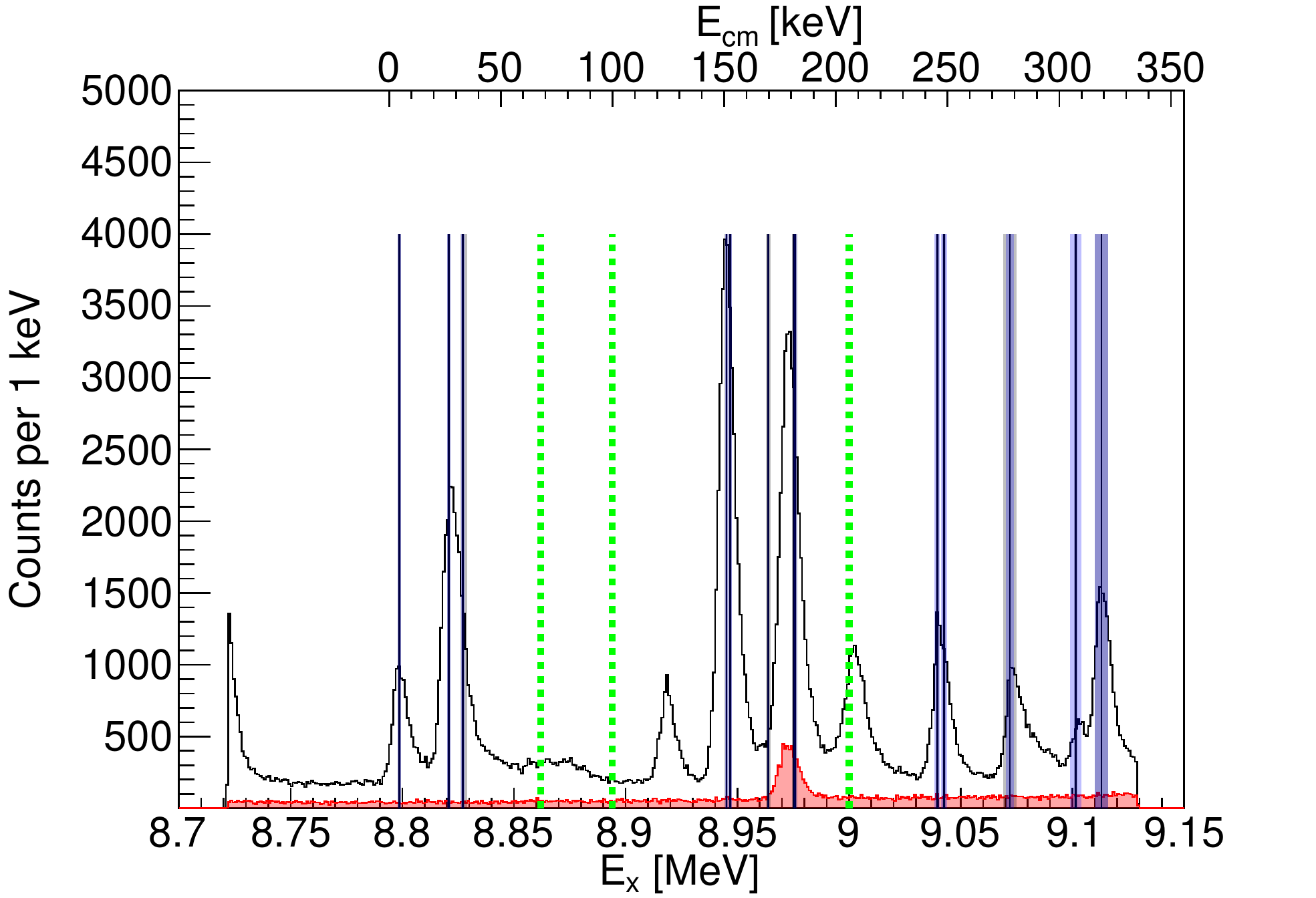}
    \includegraphics[width=0.9\columnwidth]{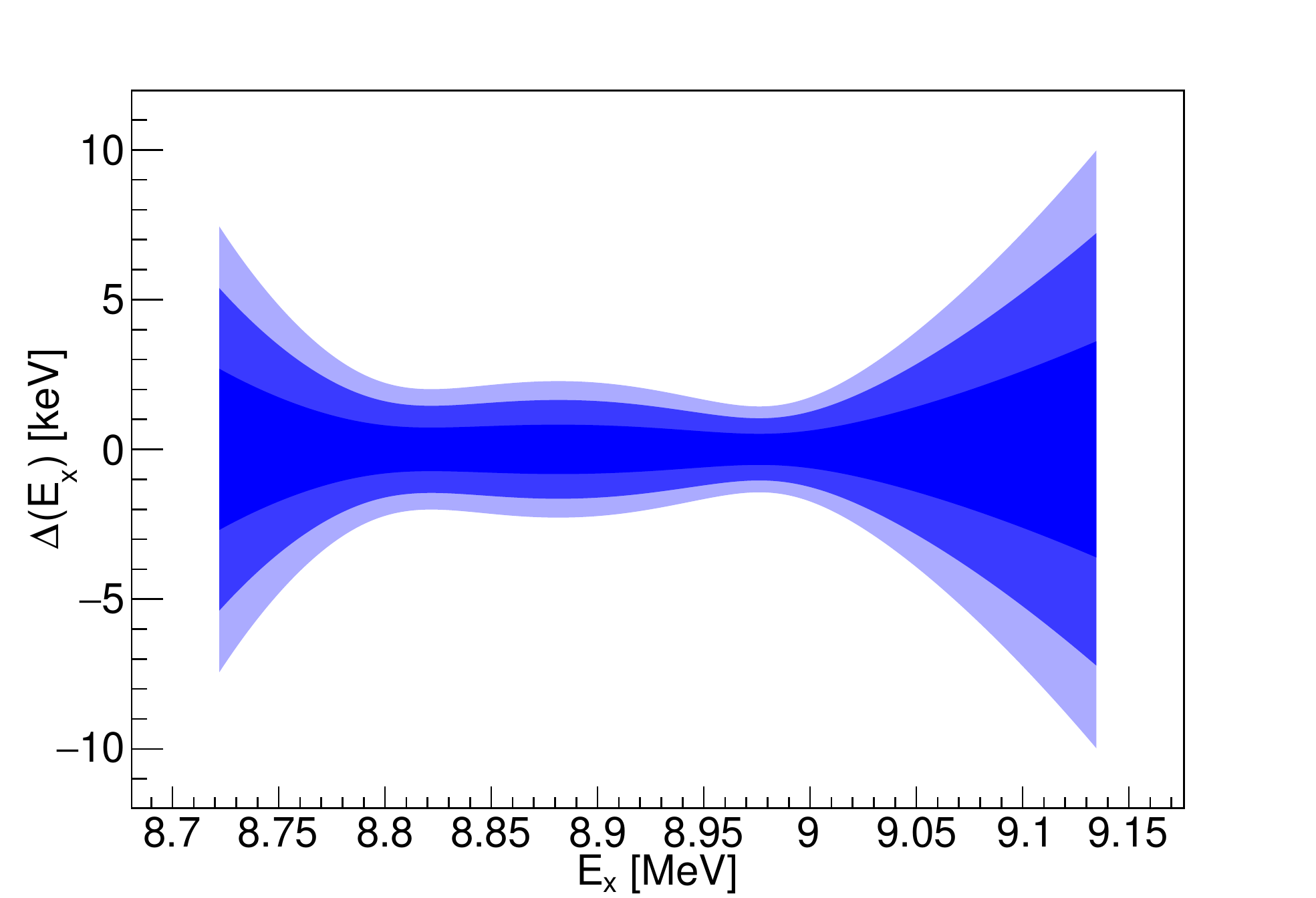}
    \caption{Same as Fig. \ref{fig:50deg} at $\theta_\mathrm{Q3D} = 35$ degrees. The $^{23}$Na states at $E_\mathrm{x} = 8798.7(8)$ and $9113(3)$ keV, the $^{19}$F state at $E_\mathrm{x} = 8864(4)$ keV and the $^{16}$O state at $E_\mathrm{x} = 8871.9(5)$ keV were used to calibrate at this angle.}
    \label{fig:35deg}
\end{figure}

\begin{figure}[t]
    \centering
    \includegraphics[width=0.9\columnwidth]{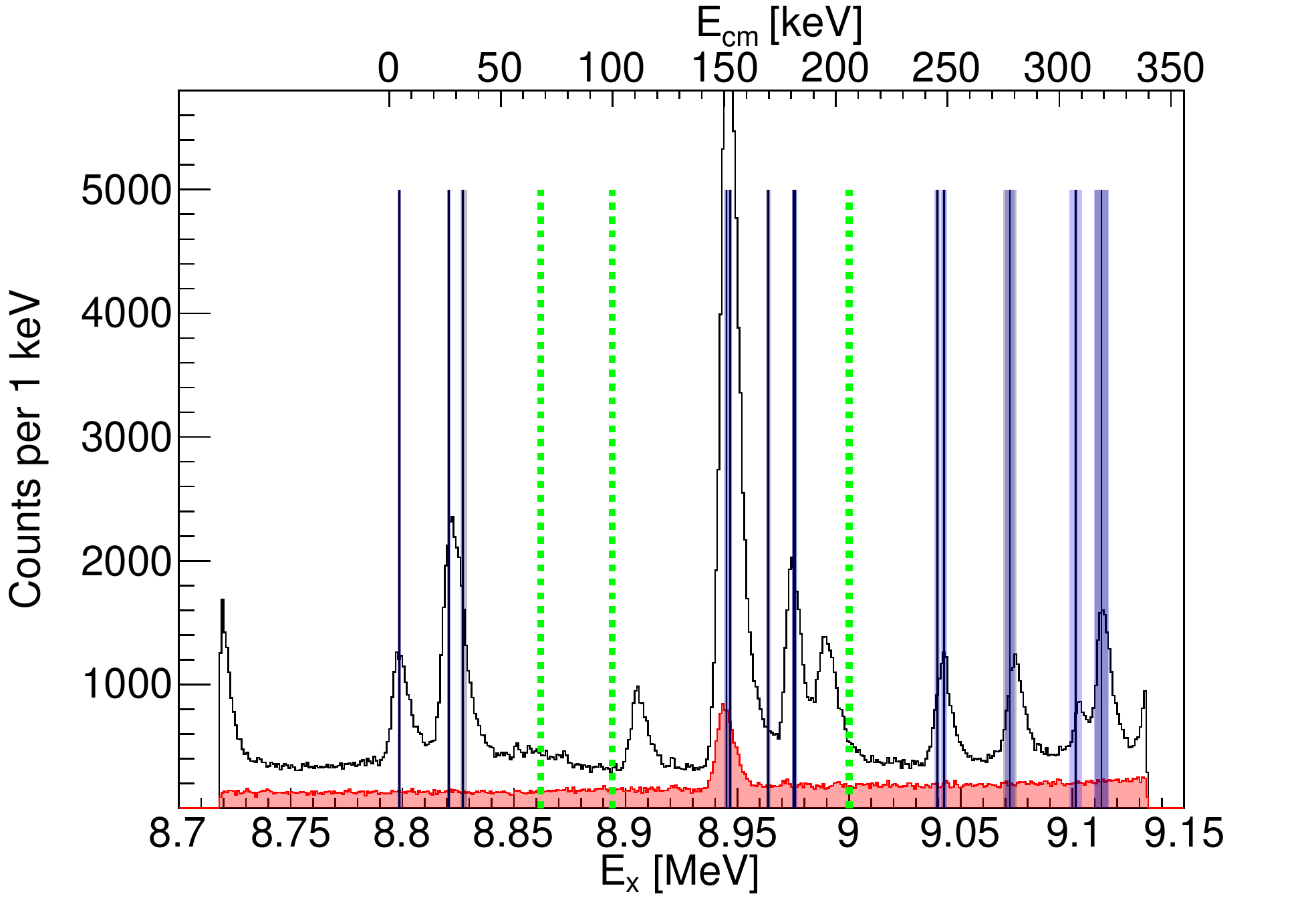}
    \includegraphics[width=0.9\columnwidth]{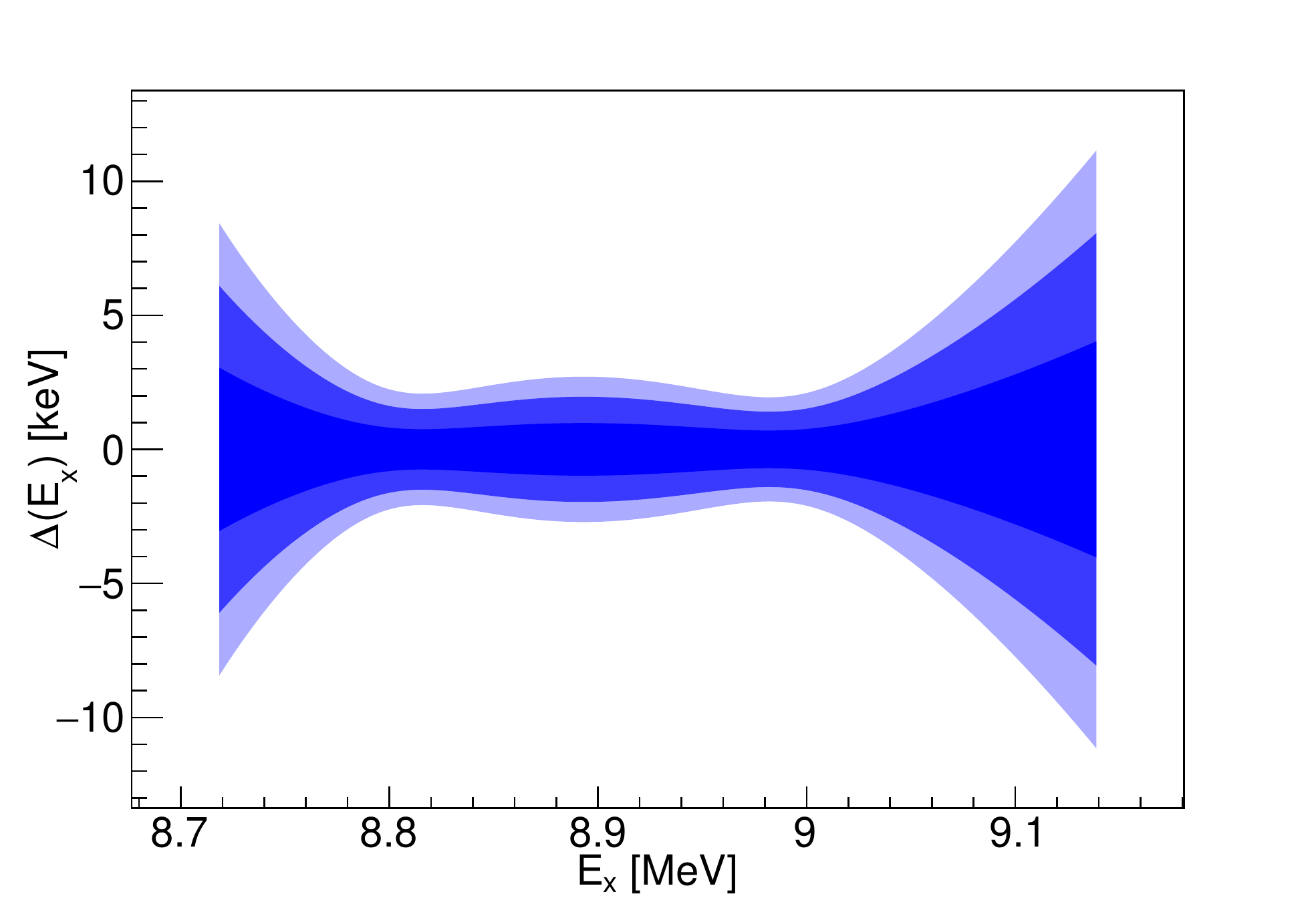}
    \caption{Same as Fig. \ref{fig:50deg} but for $\theta_\mathrm{Q3D} = 25$ degrees. The $^{23}$Na states at $E_\mathrm{x} = 8798.7(8)$, $8975.3(7)$ and $9113(3)$ keV and the $^{19}$F state at $E_\mathrm{x} = 8864(4)$ keV were used to calibrate at this angle.}
    \label{fig:25deg}
\end{figure}

The focal-plane position was related to the magnetic rigidity using known states in $^{16}$O, $^{19}$F and $^{23}$Na for calibration. For the 70-degree data, the focal-plane to rigidity calibration for these states is shown in Fig. \ref{fig:70degcalibration}. The corresponding $^{23}$Na excitation-energy spectrum is shown in Fig. \ref{fig:70deg} along with the $3\sigma$ uncertainty bands for the excitation energy. Similar calibrations to $^{23}$Na, $^{19}$F and $^{16}$O states were performed for other angles. The resulting $E_\mathrm{x}$ spectra and uncertainty bands are shown in Figs. \ref{fig:50deg}-\ref{fig:25deg}. Each of the calibrations included at least one state from a different nucleus ($^{19}$F from the NaF targets and oxygen contamination in the carbon targets to verify the assigned $^{23}$Na levels in the calibrations). Information on the states used to calibrate at each angles is given in the figure captions.

The increase in the excitation-energy uncertainty at higher excitation energies is due to the increased uncertainty in the excitation energies of the states used in the calibration at the low-rigidity (high-excitation energy) end of the focal plane. At these higher excitation energies, the $E_\mathrm{x} = 9072(3)$-keV state in $^{23}$Na has been used for the calibration when possible. However, since the excitation energy of this state has a $3$-keV uncertainty, the corresponding uncertainty band in the present experiment is of the same size.


\section{Discussion}
\label{sec:results}

There are three tentative resonance states listed in this excitation-energy region in $^{23}$Na: at $E_\mathrm{x} = 8862$, $8894$ and $9000$ keV, note that no uncertainties were given in the original study which claimed these tentative resonances. Around the tentative $E_\mathrm{x} = 8862$- and $8894$-keV states the spectrum is notably featureless. Both of the tentative states are well resolved from other known $^{23}$Na levels. It is unlikely that these resonance states exist.

The $E_\mathrm{x} = 9000$-keV state is located in a region with multiple populated levels, including contaminant states from other nuclei including $^{19}$F. Since there is no state populated at multiple angles with consistent excitation energy, we also conclude that this state has not been populated in this experiment and that it is unlikely to exist.

In this experiment, we have not observed a state at $E_\mathrm{x} = 8963.9(11)$ keV which has been assigned in $\gamma$-ray decays following the $^{12}$C($^{12}$C$,p$)$^{23}$Na fusion-evaporation reaction in coincidence with the $E_\gamma = 440$-keV transition from the decay of the first excited state in $^{23}$Na \cite{PhysRevC.87.064301,PhysRevC.30.527}. It is possible that the $\gamma$-ray is feeding the $J^\pi = 7/2^+$ state at $E_\mathrm{x} = 2075.9(4)$ keV and subsequent the $E_\gamma = 1636$-keV transition was not observed and so the excitation energy of this state has been misplaced in those $\gamma$-ray studies, though this is not confirmed  \cite{PrivateCommDave}. Alternatively, if the state is weakly populated in the present experiment then its close proximity to the $E_\mathrm{x} = 8946.8(6)$-keV may make it hard to observe.


\section{Conclusions}
\label{sec:conclusions}
The major remaining uncertainty in the $^{22}$Ne($p,\gamma$)$^{23}$Na reaction rate contributing to the Na-O anticorrelation observed in globular clusters is the existence of a state at $E_\mathrm{x} = 8862$ keV, corresponding to a resonance energy of $E_\mathrm{r,c.m.} = 68$ keV. A strong contribution of the potential $E_\mathrm{r,c.m.} = 100$-keV and $E_\mathrm{c.m.} = 206$-keV resonances (from purported states at $E_\mathrm{x} = 8894$ and $9000$ keV) to the reaction rate had previously been ruled out by previous direct measurements which provide a tight upper limit on the resonance strength of these resonances (see e.g. Ref. \cite{PhysRevLett.121.172701}). The influence of the $E_\mathrm{r,c.m.} = 68$-keV resonance on the reaction rate remained an open question largely due to the lower resonance energy. In the present measurement, which used the non-selective $^{23}$Na($p,p^\prime$)$^{23}$Na reaction to populate excited states in $^{23}$Na, none of the potential resonance states are observed, in agreement with other direct and indirect studies of the excited levels of $^{23}$Na. Only one study reports these tentative resonance states \cite{PhysRevC.4.2030} and multiple studies using different experimental probes \cite{PhysRevC.65.015801,PhysRevLett.115.252501,PhysRevLett.121.172701} including the present one see no evidence for them. These results strongly indicate that the potential resonance states do not exist (though it is not possible to prove that) and should be omitted from evaluations of the reaction rates until and unless conclusive evidence may be obtained for the existence of these states. Efforts to directly measure these resonances are unlikely to observe any signal above the direct-capture contribution. Finally, we recommend using the reaction-rate evaluation of Ref. \cite{PhysRevC.102.035801} and its associated uncertainties for future astrophysical models, reducing the reaction-rate uncertainty in the temperature range of hot-bottom burning from an order of magnitude to 40\%.

\begin{acknowledgments}

The authors thank the beam operators at the Maier-Leibnitz Laboratorium for their work and support and the target preparation laboratory of INFN-LNS for sample and target preparation and characterisation. DPCR thanks the Department of Energy Nuclear Physics grant number DE-FG02-93ER40773 and DE-SC0022469, the Texas Research Expanding Nuclear Diversity (TREND) program, the Cyclotron Institute at Texas A\&M University. MW acknowledges support provided by the Natural Sciences \& Engineering Research Council of Canada grant SAPPJ-2019-00039. PA thanks the Claude Leon Foundation for support during his time at the University of the Witwatersrand and iThemba LABS in the form of a postdoctoral fellowship. PA also thanks David Jenkins of the University of York for useful comments relating to the $^{12}$C($^{12}\mathrm{C},p$)$^{23}$Na Gammasphere data and Richard Longland of North Carolina State University for useful comments and references regarding $^{22}$Ne($p,\gamma$)$^{23}$Na direct measurements.

\end{acknowledgments}

\newpage

\bibliography{refs}

\end{document}